\documentclass{article}
\usepackage{spconf,amsmath,graphicx}

\usepackage{amsmath}
\usepackage{nccmath}
\usepackage{cite}
\usepackage{amssymb}
\newcommand{\blskip}{\baselineskip   10.20pt}
\newcommand{\ftblskip}{\baselineskip 10.20pt}
\def\0{{\mathbf{0}}}
\def\pacou{p_{\mathtt{acou}}}
\def\Plang{P_{\mathtt{lang}}}
\def\Presc{P_{\mathtt{resc}}}

\def\hatbfw{\hat{\mathbf{w}}}
\title{LATTICE RESCORING BASED ON LARGE ENSEMBLE OF\\COMPLEMENTARY NEURAL LANGUAGE MODELS}
\name{Atsunori Ogawa, Naohiro Tawara, Marc Delcroix, and Shoko Araki}
\address{NTT Corporation, Japan}
\begin{document}
\ninept
\blskip
\maketitle
\begin{abstract}
We investigate the effectiveness of
using
a large ensemble of
advanced neural language models (NLMs)
%on lattice rescoring
%for error reduction
%of given
for lattice rescoring on
automatic speech recognition (ASR) hypotheses.
Previous studies
have reported
the effectiveness of combining
%a few NLMs.
a small number of NLMs.
In contrast,
%In this study,
%However,
in this study,
we combine up to eight NLMs,
%We combine up to eight NLMs,
%,
i.e.,
%that are
%
%which are
forward/backward
long short-term memory/Transformer-LMs
%(LSTM)
%which are
that are
trained
with two different random initialization seeds.
%,
%i.e.,
%eight NLMs in total.
%
We combine these NLMs
through
%with
iterative lattice generation.
Since these NLMs
work complementarily with each other,
by combining them one by one at each rescoring iteration,
language scores attached
to
given
lattice arcs
can be
gradually refined.
%and,
%consequently,
Consequently,
errors
of
the
ASR hypotheses
can be
gradually reduced.
We also investigate the effectiveness of
%context (previous rescoring results)
%carry-over
%for
carrying over
contextual information
(previous rescoring results)
across
a lattice sequence of
%a long (e.g., lecture) speech.
a long speech
%recording
such as a lecture speech.
%a long speech (e.g., a lecture speech).
%along with the large ensemble of NLMs.
%
In experiments using a lecture speech corpus,
by combining the eight NLMs
and using context carry-over,
we obtained
a $24.4\%$ relative
word error rate
reduction from the ASR 1-best baseline.
For further comparison,
we performed
simultaneous (i.e., non-iterative) NLM combination and
100-best rescoring
using
the large ensemble of NLMs,
%,
%and
which
confirmed the advantage of lattice rescoring
with iterative NLM combination.
%over $N$-best rescoring.
%
\end{abstract}
%
%%%%%%%%%%%%%%%%%%%%%%%%%%%%%%%%%%%%%%%%%%%%%%%%%%%%%%%%%%%%%%%%%%%%%%%%%%%%%%%
%\vspace{-2.50mm}
\vspace{-3.50mm}
\section{Introduction}
\label{sec_intro}
%\vspace{-0.50mm}
\vspace{-1.50mm}
%%%%%%%%%%%%%%%%%%%%%%%%%%%%%%%%%%%%%%%%%%%%%%%%%%%%%%%%%%%%%%%%%%%%%%%%%%%%%%%
%
Based on the
recent
introduction
%rapid progress
of state-of-the-art neural network (NN) modeling,
%\cite{Hinton_IEEESPM2012,Yu_Springer2015},
%recently,
%The performance
%the accuracy
the performance
of automatic speech recognition (ASR)
has
been
%recently
greatly improved
\cite{Hinton_IEEESPM2012,Yu_Springer2015},
and various types of
ASR-based
%ASR
applications,
including
%such as
voice search services and smart speakers,
have been actively developed.
Despite this great progress,
in
some
situations such as
performing ASR in noisy environments
%and/or
or
performing ASR
%that
for
%casual-style
conversational
speech,
%the ASR accuracy
the accuracy of ASR
%its accuracy
remains at an unsatisfactory level
%can still be degraded
%\cite{Vincent_CSL2016,Barker_IS2018,Saon_IS2015,Saon_IS2016,Xiong_arXiv2017,Xiong_ICASSP2017,Xiong_ICASSP2018}.
%\cite{Barker_IS2018,Saon_IS2015,Saon_IS2016,Xiong_arXiv2017,Xiong_ICASSP2018}.
\cite{Saon_IS2016,Saon_IS2017,Xiong_arXiv2017,Xiong_ICASSP2018,Barker_IS2018,Watanabe_CHiME6}.

%To improve the ASR accuracy
%in such severe situations,
%a promising method is to
A promising
%method
approach
%for improving the ASR accuracy
for reducing ASR errors
%to reducing ASR errors
in such severe situations
%is to
involves
%exploit
%use
the use of
multiple
%speech recognition
ASR
hypotheses
(word sequences),
which are represented
%in such a form as
in such forms as
an $N$-best list
%and
or
a lattice.
%in the form of,
%e.g.,
%an
%$N$-best list
%and
%a
%lattice.
%,
%and
%word
%a
%confusion network
%\cite{Mangu_CSL2000}.
%
This is because
%the 1-best speech recognition hypothesis
%the 1-best ASR hypothesis
%for an utterance
%may contain many errors,
%but
%the
%a
an ASR
hypothesis that has a significantly lower word error rate (WER)
than the 1-best hypothesis
can be found in
%the
multiple
%other
hypotheses
if it is appropriately rescored
(reranked).
%
%Actually,
%various
Various
types of rescoring methods have been
%actually
developed and applied
to noisy
%and/or
or
conversational speech recognition
%,
%e.g.
%\cite{Vincent_CSL2016,Barker_IS2018,Saon_IS2015,Saon_IS2016,Xiong_arXiv2017,Xiong_ICASSP2018,Erdogan_CHiME4,Du_CHiME4,Menne_CHiME4,Du_CHiME5,Kanda_CHiME5,Zhao_CHiME5}.
\cite{Saon_IS2016,Saon_IS2017,Xiong_arXiv2017,Xiong_ICASSP2018,Du_CHiME5,Kanda_CHiME5,Zhao_CHiME5,Li_CHiME5,Medennikov_CHiME6,Arora_CHiME6,Zorila_CHiME6,Zmolikova_CHiME6,Irie_IS2018,Xiong_EMNLP2018,Irie_IS2019,Parthasarathy_arXiv2019,Irie_ASRU2019,Sun_ICASSP2021}.

In these rescoring methods,
advanced neural language models (NLMs)
%which are difficult to be used
%in the first-pass decoding
%of hybrid ASR systems,
are used
as rescoring models.
%[88-88].
%
%since they
They
can
accurately
%capture
model
%estimate probabilities of
%greatly
much
longer word sequences
%(contexts)
%a greatly
%longer word sequence
%(context)
%than
%compared with
than can
%traditional
conventional
count-based $n$-gram LMs
\cite{Jelinek_MITPress1998,Kneser_ICASSP1995},
which can
%that can
%estimate probabilities of
model
%a sequence of only $n$ words
sequences of only $n$ words
(where $n$ is typically three to five).
%($n$ is typically three to five).
%
%They are used
These NLMs are used
to refine language scores
attached to ASR hypotheses
that
are calculated using
the $n$-gram LMs.
%and attached to ASR hypotheses.
%
Among the
%neural LMs,
NLMs,
%currently,
long short-term memory (LSTM)-based recurrent NLMs
\cite{Sundermeyer_IS2012}
are
currently
the most widely used
%type.
model.
%as rescoring models
%in rescoring
%[88-88].
%
A forward LSTMLM can provide good WER reduction,
%[88-88]
but
%it
the WER
can be further reduced 
by additionally using another model.
%e.g.,
%for example,
Such a model
%might be,
would be,
%is,
for example,
a forward LSTMLM that has a different model
%size
structure
\cite{Li_CHiME5,Medennikov_CHiME6}
%,
%a forward LSTMLM
%that is
or that is
%trained with a different initialization seed
trained with a different setting
%(e.g., a different initialization seed and a different data shuffling scheme)
(e.g., a different initialization seed or a different data shuffling scheme)
%(e.g., a random initialization seed and a data shuffling scheme)
%(e.g., using a random initialization seed and a data shuffling scheme)
\cite{Xiong_arXiv2017,Irie_IS2018}
%,
%and
or
a backward LSTMLM
that is
trained by using
%a
%reversed text data
a reversed text dataset
\cite{Xiong_arXiv2017,Xiong_ICASSP2018,Irie_IS2018,Kanda_CHiME5,Arora_CHiME6},
%the WER can be further 
since these
%LSTMLMs
models
work complementarily with each other.
%
%%%%%%As described above,
%%%%%%an
%%%%%An $n$-gram LM cannot model long contexts
%%%%%as described above,
%%%%%%
%%%%%%However,
%%%%%but,
%%%%%in other words,
%%%%%if focuses on modeling local contexts
%%%%%in contrast to an LSTMLM,
%%%%%and thus it 
%
In addition to the LSTMLMs,
%more recently,
%Transformer-based LMs
%\cite{Vaswani_NIPS2017}
NLMs based on Transformers
\cite{Vaswani_NIPS2017}
have recently been used
%as rescoring models
%in rescoring.
for rescoring.
%\cite{Vaswani_NIPS2017}.
%
They 
have a non-recurrent
%model
%self-attention
self-attentive
architecture
%,
%which
that
is completely different from that of the LSTMLMs,
and
they
show comparable or superior
rescoring performance to the LSTMLMs
\cite{Irie_IS2019,Irie_ASRU2019,Sun_ICASSP2021}.

%In a long speech
%(a series of utterances)
%%recording
%such as a lecture speech,
%the content of an utterance is naturally influenced
%by the contexts of the previous utterances.
%
By performing ASR for a long speech
%recording
%(a series of utterances)
such as a lecture speech,
a long ASR hypothesis sequence
%,
%which corresponds to a series of utterances,
%is obtained.
can be obtained.
In such a long speech
(a series of utterances),
the content of an utterance is naturally influenced
%by the contents of the previous utterances (context).
by the content of previous utterances (i.e., context).
Therefore,
%for rescoring
in rescoring
such a
long ASR hypothesis sequence,
%of the ASR hypothesis sequence of such a long speech,
%in addition to use the above described advanced neural LMs,
it is reasonable
to use the rescoring results of the previous hypotheses
as
contextual information
%the context
%for the current hypothesis rescoring.
for rescoring the current hypothesis.
%
%Actually,
%it is reported that,
%It is reported that,
It has been reported that,
%%%in addition to use
%%%the
%%%advanced neural LMs,
by
%using
carrying over
contextual information
%rescoring results
across ASR hypotheses,
%using the neural LMs,
the rescoring performance
for such a long ASR hypothesis sequence
can be improved
\cite{Xiong_arXiv2017,Xiong_ICASSP2018,Xiong_EMNLP2018,Parthasarathy_arXiv2019,Irie_ASRU2019,Zmolikova_CHiME6,Sun_ICASSP2021}.
%can be improved
%for such a long ASR hypothesis sequence [88-88].
%can be improved [88-88].
%described above,
%the WER can be further reduced
%%%the performance of the current hypothesis rescoring
%%%can be further improved
%%%by inheriting rescoring results of the previous hypotheses 
%%%as contextual information [88-88].
%in the current hypothesis rescoring [88-88].
%
%Since this ASR hypothesis sequence
%corresponds to a series of utterances
%
%Since 
%the content of an utterance is naturally influenced
%by the contexts of the previous utterances in such a long speech,

%%%%%In this study,
%%%%%we focus on lattice rescoring
%%%%%and propose a lattice rescoring method
%%%%%that is based on
%%%%%%bidirectional
%%%%%%(both forward and backward)
%%%%%iterative lattice generation.
%%%%%%
%%%%%We use
%%%%%forward/backward LSTM/Transformer-based LMs one by one
%%%%%at each rescoring iteration to gradually refine language scores
%%%%%attached to given lattice arcs
%%%%%%with expecting their complementarity
%%%%%based on their complementarity
%%%%%(Section~\ref{ssec_gen}).
%
In this study,
we investigate the effectiveness of using a large ensemble of NLMs
on lattice rescoring.
As described above,
previous studies
\cite{Saon_IS2016,Saon_IS2017,Li_CHiME5,Medennikov_CHiME6,Arora_CHiME6}
have reported the effectiveness of combining
%a few NLMs
%(up to four NLMs \cite{Medennikov_CHiME6})
%a few
a small number of NLMs
(up to four \cite{Medennikov_CHiME6})
%NLMs
on lattice rescoring.
%in lattice rescoring.
%\cite{Saon_IS2016,Saon_IS2017,Li_CHiME5,Medennikov_CHiME6,Arora_CHiME6}.
%\cite{Saon_IS2016,Saon_IS2017,Xiong_arXiv2017,Xiong_ICASSP2018,Kanda_CHiME5,Li_CHiME5,Medennikov_CHiME6,Arora_CHiME6,Irie_IS2018,Sun_ICASSP2021}.
%
In contrast,
we combine up to eight NLMs,
i.e.,
forward/backward LSTM/Transformer-LMs,
%that are trained with two different random initialization seeds.
which are trained with two different random initialization seeds.
We combine these complementary NLMs through iterative lattice generation
%(Section~\ref{ssec_ilg})
while introducing context carry-over
%(Section~\ref{ssec_cco}).
%\cite{Xiong_arXiv2017,Xiong_ICASSP2018,Xiong_EMNLP2018,Parthasarathy_arXiv2019,Irie_ASRU2019,Zmolikova_CHiME6,Sun_ICASSP2021}
(Section~\ref{sec_lr}).
We conducted experiments
including
%experimental settings that were not investigated
experimental settings that have not
%yet
been investigated
in previous studies
(Section~\ref{sec_rel})
%.
%
%In experiments using a lecture speech corpus,
%we show the effectiveness of the large ensemble of NLMs on lattice rescoring
%(Section~\ref{sec_exp}).
and confirmed
the effectiveness of
%the
using a
large ensemble of
%the
NLMs
%on
for
lattice rescoring
(Section~\ref{sec_exp}).
%
%We believe
%that our experimental results are very informative in the research area
%of ASR hypothesis rescoring.
%
%We can summarize our
%main findings as follows.
Our main findings can be summarized
as follows.
\renewcommand{\labelenumi}{(\theenumi)}
\begin{enumerate}
\item Combining
six or seven NLMs can improve
%the lattice rescoring performance.
the performance of lattice rescoring.
\item Lattice rescoring has an advantage over $N$-best rescoring
%to
%handle
%exploit
%the
%for exploiting
%when exploiting
when using
a large ensemble of NLMs.
\item Performing context carry-over in the backward direction is as effective
as performing it in the forward direction.
\item Iterative NLM combination has the potential to outperform simultaneous
NLM combination,
%in the fast lattice rescoring setting.
%in a fast lattice rescoring setting.
%in the fast rescoring setting.
especially in a fast lattice rescoring setting.
\end{enumerate}
%
%%%%%%%%%%%%%%%%%%%%%%%%%%%%%%%%%%%%%%%%%%%%%%%%%%%%%%%%%%%%%%%%%%%%%%%%%%%%%%%
\vspace{-2.00mm}
\section{Lattice rescoring method}
\label{sec_lr}
\vspace{-1.00mm}
%%%%%%%%%%%%%%%%%%%%%%%%%%%%%%%%%%%%%%%%%%%%%%%%%%%%%%%%%%%%%%%%%%%%%%%%%%%%%%%
%
We
%describe
introduce
a
%simple
%framework
method
for combining NLMs
through
%with
iterative lattice generation
and
a
%framework
method
for carrying over
contextual information
across lattices.
%
%******************************************************************************
%\vspace{-1.00mm}
\subsection{Combining NLMs through iterative lattice generation}
\label{ssec_iter}
%\vspace{-0.00mm}
%******************************************************************************
%
A lattice is an efficient ASR result form of an input utterance
that includes multiple ASR
hypothesis candidates
%hypotheses
of the utterance.
A lattice consists of nodes and arcs,
where
a node corresponds to a word boundary
%and
while
an arc corresponds to a recognized word.
An arc 
has an acoustic score
%(log likelihood)
and a language score,
%and it
%has acoustic and language scores
%(log probability),
which
%that
are calculated during
%the first-pass
the 
%ASR 1st-pass decoding.
ASR first-pass decoding.
Language scores
are usually calculated using
%an $n$-gram LM.
a count-based $n$-gram LM.

We use the push-forward algorithm
\cite{Auli_EMNLP2013,Sundermeyer_IEEEACMTASLP2015,Liu_IEEEACMTASLP2016,Kumar_ASRU2017}
for lattice rescoring.
%
%In lattice rescoring,
Given a lattice for an
input
utterance,
%using a rescoring model,
%(a neural LM),
we perform search
on the lattice
%from the lattice begin node
from its begin node
%based on the push-forward algorithm [88-88]
to refine the language scores attached to the
%given
%lattice
arcs
using a rescoring model.
Then,
by tracing back the
%refined lattice
rescored lattice
from its end node,
we can obtain
the final ASR hypothesis, i.e.,
the best word (arc) sequence,
%that has the highest score.
that shows the highest score.
%to refine

\begin{figure}[t]
  \centering
  \includegraphics[width=0.770\linewidth]{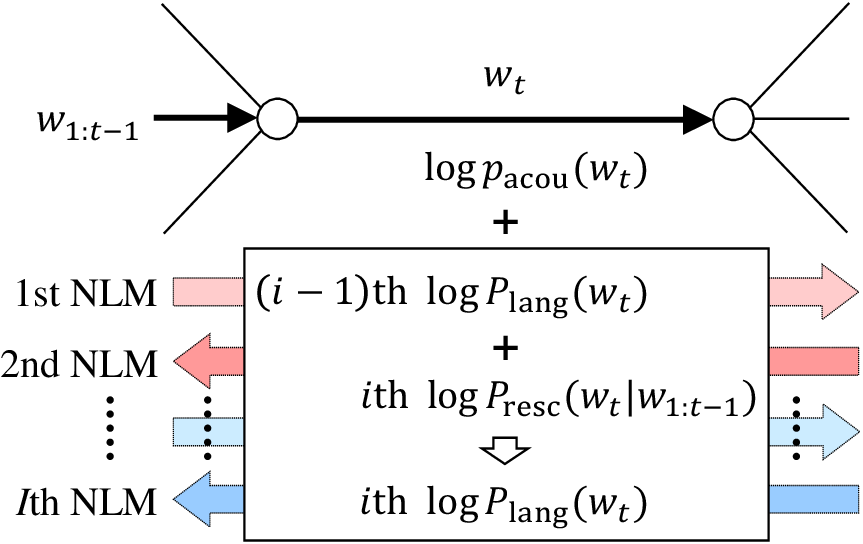}
  \vspace{-1.50mm}
  \caption{\ftblskip Iterative lattice generation (language score refinement) at a lattice arc using $I$ complementary NLMs.}
  \label{fig_iter}
  \vspace{5.25mm}
%\end{figure}
%------------------------------------------------------------------------------
%\begin{figure}[t]
  \centering
  \includegraphics[width=1.000\linewidth]{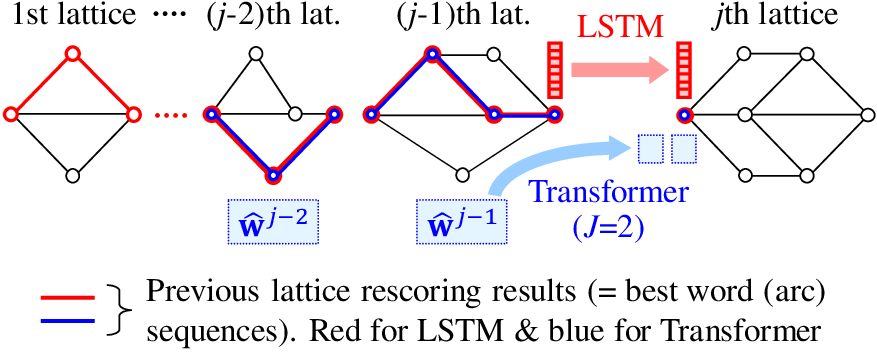}
  \vspace{-6.00mm}
  %\caption{\ftblskip Framework for carrying over contextual information across lattices (utterances).}
  %\caption{\ftblskip Methods for carrying over contextual information (previous rescoring results) across lattices (utterances) with LSTM/Transformer-LMs.}
  %\caption{\ftblskip Carrying over contextual information (previous rescoring results) across lattices (utterances) with LSTM/Transformer-LMs.}
  \caption{\ftblskip Methods for carrying over contextual information (previous rescoring results) across lattices (utterances) with NLMs.}
  \label{fig_cco}
  \vspace{-2.25mm}
\end{figure}

We focus on the search processing at
%an arc
a lattice arc
as
shown in Fig.~\ref{fig_iter}.
Let $w_{1:t-1}$ be
a partial word (arc) sequence
(hypothesis)
%a partial hypothesis
%(word (arc) sequence)
%(hypothesis)
of length $t\!-\!1$.
It is extended
from the lattice begin node
and its current score (log-likelihood)
is $\log{p(w_{1:t-1})}$.
%
%It reaches the arc $w_{t}$ that has
%the acoustic score (log-likelihood)
%$\log{\pacou(w_{t})}$
It reaches the arc $w_{t}$, which has
an acoustic score (log-likelihood)
$\log{\pacou(w_{t})}$
and
a language score (log probability)
$\log{\Plang(w_{t})}$.
By extending the partial hypothesis $w_{1:t-1}$ to this arc,
the score of the extended partial hypothesis $w_{1:t}$ can be obtained as,
\begin{fleqn}
\begin{multline}
%\begin{equation}
\log{p(w_{1:t})}
=\log{p(w_{1:t-1})}
+\log{\pacou(w_{t})} \\
+\alpha\left\{\underline{(1-\beta)\log{\Plang(w_{t})}
+\beta\log{\Presc(w_{t}\,|\,w_{1:t-1})}}\right\},
\label{eq_resc}
%\end{equation}
\end{multline}
\end{fleqn}
%
%\begin{fleqn}
%\begin{multline}
%\log{p^{i}(w_{1:t})}
%=\log{p^{i}(w_{1:t-1})}
%+\log{\pacou(w_{t})} \\
%+\alpha\left\{\underline{(1-\beta^{i})\log{\Plang^{i}(w_{t})}
%+\beta^{i}\log{\Presc^{i}(w_{t}|w_{1:t-1})}}\right\}
%\end{multline}
%\end{fleqn}
%
where
$\log{\Presc(w_{t}|w_{1:t-1})}$
is the language score of $w_{t}$ given $w_{1:t-1}$
calculated using a rescoring model (an NLM),
$\beta$ $(0\!<\!\beta\!<\!1)$ is the interpolation weight
between the original language score
and that calculated
%with
using
the rescoring model,
and $\alpha$ $(\alpha\!>\!0)$ is the weight of the language score
against the acoustic score.
The underlined
term
%part
in Eq.~(\ref{eq_resc})
corresponds to 
%is
the refined language score attached to the arc $w_{t}$.
By performing
%the above-described
this
search processing at all the arcs
in the given lattice,
we can generate a
%refined lattice
rescored lattice
from the original lattice.
%
%Note that,
%depending on the hyperparameters of search
%(e.g., $n$ of
%$n$-gram approximation for merging hypotheses at a node
%and the maximum number of hypotheses ($k$)
%stored
%at a node),
%the structure of the refined lattice can be changed from
%that of the original lattice
%\cite{Sundermeyer_IEEEACMTASLP2015,Liu_IEEEACMTASLP2016,Kumar_ASRU2017,Xu_ICASSP2018}.
%
%Note that,
%depending
Depending on the hyperparameters of search,
the structure of the
%refined lattice
rescored lattice
%can be changed from
can change from
that of the original lattice
\cite{Sundermeyer_IEEEACMTASLP2015,Liu_IEEEACMTASLP2016,Kumar_ASRU2017,Xu_ICASSP2018}.
The hyperparameters
%are,
include,
for example,
the $n$ of
$n$-gram approximation for merging hypotheses at a node
%(hypotheses that have the same word history of the length no more than $n$
%words are merged into one hypothesis)
%(hypotheses that have the same history of no more than $n$ words
%are merged into one hypothesis)
%(merging hypotheses that have the same history of no more than $n$ words
(merging hypotheses that have the same history of $n$ or more words
into a single hypothesis)
%at a node)
and the maximum number of hypotheses ($k$)
stored at a node.

In the lattice generation described above,
we performed processing in the forward direction
using a forward NLM as the rescoring model.
As described in Section~\ref{sec_intro},
it has been reported that,
%by performing a few more processings using other NLMs
by combining a few more NLMs
that have
%the
complementarity with the above forward NLM,
we can obtain a
%large
steady
WER reduction
%\cite{Saon_IS2016,Saon_IS2017,Xiong_arXiv2017,Xiong_ICASSP2018,Kanda_CHiME5,Li_CHiME5,Medennikov_CHiME6,Arora_CHiME6,Irie_IS2018,Sun_ICASSP2021}.
\cite{Saon_IS2016,Saon_IS2017,Li_CHiME5,Medennikov_CHiME6,Arora_CHiME6}.
%since these neural LMs work complementarily with each other
%by performing a backward processing using a backward neural LM
%after the forward processing,
%we can obtain a large WER reduction
%since these neural LMs work complementarily with each other
%[88-88].
%
In this study,
to further reduce the WER,
we
%continue
repeat
lattice generation
%(language score refinement)
for more iterations
%(in our experiments,
(up to eight iterations
%(eight iterations
%at maximum
in our experiments
%see Section~\ref{sec_exp})
%(Section~\ref{sec_exp}))
%see Section~4.3)
as described in Section~\ref{sec_exp})
by changing the NLMs,
which are complementary with each other,
at each iteration
%to further reduce the WER.
as shown in Fig.~\ref{fig_iter}
($I=8$).
%
%%%%%These NLMs
%%%%%have
%%%%%different processing directions
%%%%%(forward or backward),
%%%%%different architectures
%%%%%%(an LSTMLM or a Transformer-LM),
%%%%%(LSTM-based or Transformer-based),
%%%%%%(LSTM- or Transformer-based),
%%%%%and different
%%%%%%training
%%%%%%model parameter
%%%%%random
%%%%%initialization seeds
%%%%%for training
%%%%%%seeds for model parameter initialization
%%%%%to guarantee their complementarity.

To perform
iterative lattice generation
(language score refinement),
%using the NLMs,
we need to design a way to
%weight
interpolate
the
$(i\!-\!1)$th
%$(i-1)$th
language score attached to a lattice arc
and the $i$th language score calculated using
%an NLM
the NLM
%a rescoring model
(see Eq.~(\ref{eq_resc}) and Fig.~\ref{fig_iter}).
In this study,
we assume that all the NLMs
contribute equally
%for refining
%to refine
to refining
language scores
(actually,
in our experiments,
they show similar
%evaluation data
dev/eval data
perplexities as shown
in Table~\ref{tab_ppl})
and define the $i$th interpolation weight as,
%$\beta(i)=1/(1+i)$.
%
\begin{fleqn}
\begin{equation}
\beta(i)=\frac{1}{1+i}.
\label{eq_beta}
\end{equation}
\end{fleqn}
%
%By this definition,
With this definition,
at the $i$th iteration,
language scores calculated using the first to $i$th NLMs
can be equally combined at a lattice arc.
%
%Note that,
%with this definition,
Following this definition,
we combine the $n$-gram LM score
%($0$th language score)
attached to an arc in the original lattice
(the $0$th language score)
equally with
the other NLM scores.
%equally among
%the NLM scores.
%
%This is because an $n$-gram LM cannot model long contexts
%but, in other words,
%it focuses on modeling local contexts in contrast to the neural LMs,
This is because an $n$-gram LM 
focuses on modeling local contexts,
in contrast to the NLMs,
and thus it works complementarily with the NLMs
\cite{Saon_IS2016,Saon_IS2017,Xiong_arXiv2017,Xiong_ICASSP2018,Du_CHiME5,Kanda_CHiME5,Zhao_CHiME5,Li_CHiME5,Medennikov_CHiME6,Arora_CHiME6,Zorila_CHiME6,Zmolikova_CHiME6,Irie_IS2018,Xiong_EMNLP2018,Irie_IS2019,Parthasarathy_arXiv2019,Irie_ASRU2019,Sun_ICASSP2021}.
We can obtain the final ASR hypothesis
by tracing back the $I$th rescored lattice.
%
%------------------------------------------------------------------------------
%------------------------------------------------------------------------------
%
%******************************************************************************
\vspace{-1.50mm}
\subsection{Carrying over contextual information across lattices}
\label{ssec_cco}
%******************************************************************************
%
As described in Section~\ref{sec_intro},
in rescoring an ASR lattice sequence of a long speech,
%recording,
we can exploit the previous
%In case of using the previous
%(1 to $(j\!-\!1)$th)
lattice rescoring results
%for the current
%%($j$th)
%lattice rescoring
to rescore the current lattice
\cite{Xiong_arXiv2017,Xiong_ICASSP2018,Xiong_EMNLP2018,Parthasarathy_arXiv2019,Irie_ASRU2019,Zmolikova_CHiME6,Sun_ICASSP2021}.
In this case,
we need to develop a
%framework
method
to carry over
the previous rescoring results
%(the underlined part of Eq.~(\ref{eq_cco}))
(contextual information)
to the begin node of the current lattice.

When we use an LSTMLM as the rescoring model,
this
context carry-over
can be easily
%achieved.
implemented.
Thanks to its recurrent model architecture,
the LSTMLM can
%represent
encode
%all
%the previous
rescoring results
for all the previous lattices (utterances)
%with
in
a single hidden state vector.
As shown in Fig.~\ref{fig_cco},
we can copy the single hidden state vector
%at
from
the end node of the
%previous
last
($(j\!-\!1)$th)
lattice to the
begin node of the current ($j$th) lattice and use it
as the initial hidden state vector to start rescoring on the current lattice.
Note that,
when we do not perform context carry-over,
we use the zero vector as the initial hidden state vector
at the begin node of every lattice.
%(utterance).

In contrast to the LSTMLM,
when we use a Transformer-LM,
%as the rescoring model,
we need to store a sequence of hidden state vectors
to represent the previous rescoring results,
since
%, in contrast to the LSTMLM,
the Transformer-LM has a non-recurrent self-attentive model architecture
\cite{Vaswani_NIPS2017}.
%
%This means that it consumes linearly increasing memory volume
%This means that it consumes the memory linearly
%increasing memory volume
%size
This means that it linearly increases memory usage
as the
%with the
%previous
rescoring results get longer
\cite{Irie_ASRU2019}.
%
%%%%%Therefore,
%%%%%%Hense,
%%%%%in this study,
%%%%%%therefore,
%%%%%we shrink the previous rescoring results as,
%%%%%%
%%%%%\begin{fleqn}
%%%%%\begin{equation}
%%%%%v_{1:T_{1}}^{1},\cdots,v_{1:T_{j-1}}^{j-1}
%%%%%\,\rightarrow\,\,\,
%%%%%v_{1:T_{j-J}}^{j-J},\cdots,v_{1:T_{j-1}}^{j-1},
%%%%%\label{eq_shrink}
%%%%%\end{equation}
%%%%%\end{fleqn}
%%%%%%
%%%%%and keep only
%%%%%the last $J$ ($(j\!-\!J)$th to $(j\!-\!1)$th) rescoring results.
%%%%%%(the right term).
%%%%%%
%%%%%As shown in Fig.~\ref{fig_cco},
%%%%%we copy these shrunk rescoring results
%%%%%%(contexts)
%%%%%from the end node of the previous lattice
%%%%%to the begin node of the current lattice
%%%%%%as shown in Fig.~\ref{fig_cco}
%%%%%and start rescoring on the current lattice using the copied
%%%%%previous rescoring results as the context of length $J$.
%
Therefore,
in this study,
as shown in Fig.~\ref{fig_cco},
we keep only the last $J$ rescoring results,
%i.e., $c^{j-J},\cdots,c^{j-1}$,
i.e., $\hatbfw^{j-J},\cdots,\hatbfw^{j-1}$,
%i.e., $\hatbfw^{j\!-\!J},\cdots,\hatbfw^{j\!-\!1}$,
for the last $J$ lattices
and use them as contextual information
to rescore the current ($j$th) lattice.
%
%Note that,
%in the above,
In the above methods,
we perform context carry-over in the forward direction
(from the first utterance to the last utterance
in a given long speech)
using a forward NLM,
%but,
however,
similarly,
we can also perform
%context carry-over
it
in the backward direction
%(from the last utterance to the first utterance)
(from the last utterance to the first utterance)
using a backward NLM.
%
%%%%%%%%%%%%%%%%%%%%%%%%%%%%%%%%%%%%%%%%%%%%%%%%%%%%%%%%%%%%%%%%%%%%%%%%%%%%%%%
\vspace{-2.50mm}
\subsection{Other rescoring methods}
\label{ssec_others}
\vspace{-1.00mm}
%%%%%%%%%%%%%%%%%%%%%%%%%%%%%%%%%%%%%%%%%%%%%%%%%%%%%%%%%%%%%%%%%%%%%%%%%%%%%%%
%
We introduced a
%framework
method
for combining NLMs
through iterative lattice generation
in Section~\ref{ssec_iter}.
%,
%
%%%%%This
%%%%%%framework
%%%%%method
%%%%%is simple and easy to implement,
%%%%%%but
%%%%%however,
%%%%%we do not necessarily need to combine the NLMs iteratively.
%but,
%however,
However,
we do not necessarily need to combine the NLMs iteratively.
%
%Actually,
%we can combine the NLMs
In fact,
we can combine the NLMs
%We can combine the NLMs
simultaneously
%with the lattice combination technique
%with a technique like lattice combination
with, e.g., lattice combination
\cite{Li_IS2002,Xu_ICASSP2010}.
In this case,
we
individually
perform lattice rescoring using each of the NLMs
with the $n$-gram LM scores attached to arcs in the original lattice,
and
then we combine the rescored lattices with equal weight
(since,
as described in Section~\ref{ssec_iter},
we assume that all the NLMs contribute equally
%for rescoring)
to refining language scores)
%language score refinement)
while solving the differences in the rescored lattice structures.
We experimentally compare these two
%(iterative or simultaneous)
NLM combination
%frameworks
methods
in Section~\ref{sec_exp}.

$N$-best rescoring
\cite{Xiong_arXiv2017,Xiong_ICASSP2018,Kanda_CHiME5,Zmolikova_CHiME6,Irie_IS2018,Xiong_EMNLP2018,Parthasarathy_arXiv2019,Sun_ICASSP2021}
is another
%representative
widely used
ASR hypothesis rescoring method.
%than lattice rescoring.
%
As with lattice rescoring,
we can perform $N$-best rescoring using a large ensemble of NLMs.
We experimentally compare these two
%(lattice or $N$-best)
rescoring methods
in Section~\ref{sec_exp}.
Note that,
in contrast to lattice rescoring,
%In contrast to lattice rescoring,
%in $N$-best rescoring,
with $N$-best rescoring,
the iterative and simultaneous NLM combination methods
provide
%exactly
the same rescoring results.
%in $N$-best rescoring.
%
%%%%%%%%%%%%%%%%%%%%%%%%%%%%%%%%%%%%%%%%%%%%%%%%%%%%%%%%%%%%%%%%%%%%%%%%%%%%%%%
%\vspace{-1.50mm}
\vspace{-2.50mm}
\section{Relation to prior work}
\label{sec_rel}
\vspace{-1.25mm}
%%%%%%%%%%%%%%%%%%%%%%%%%%%%%%%%%%%%%%%%%%%%%%%%%%%%%%%%%%%%%%%%%%%%%%%%%%%%%%%
%
As described in Section~\ref{sec_intro},
rescoring is a promising approach
%for improving the ASR accuracy
for reducing ASR errors
and many
good
%useful
studies have been conducted on rescoring techniques
\cite{Saon_IS2016,Saon_IS2017,Xiong_arXiv2017,Xiong_ICASSP2018,Du_CHiME5,Kanda_CHiME5,Zhao_CHiME5,Li_CHiME5,Medennikov_CHiME6,Arora_CHiME6,Zorila_CHiME6,Zmolikova_CHiME6,Irie_IS2018,Xiong_EMNLP2018,Irie_IS2019,Parthasarathy_arXiv2019,Irie_ASRU2019,Sun_ICASSP2021}.
%
%%%%%In this study,
%%%%%we conduct experiments
%%%%%in Section~\ref{sec_exp}
%%%%%including
%%%%%experimental settings that were not investigated
%%%%%in previous studies
%%%%%as follows.
%
Many studies
have reported
the effectiveness of combining complementary NLMs,
but they
%try
%investigate
investigated
%the combination of a few NLMs
combinations of only a few NLMs
\cite{Saon_IS2016,Saon_IS2017,Xiong_arXiv2017,Xiong_ICASSP2018,Kanda_CHiME5,Li_CHiME5,Medennikov_CHiME6,Arora_CHiME6,Irie_IS2018,Sun_ICASSP2021}.
In contrast,
%in this study,
we combine up to eight
%complementary
NLMs.
%as described in Section~\ref{sec_exp}.
%
%In previous studies,
The previous studies
did not use
a backward Transformer-LM,
%is
%%seems to be
%not used,
but we use it as one of the
%complementary
NLMs.
The combination of an LSTMLM and a Transformer-LM
%is
was
%tried
investigated
only in \cite{Sun_ICASSP2021}
with $N$-best rescoring.
We
%evaluate
investigate
their combination
%in
with
lattice rescoring.

The effectiveness of context carry-over has also been reported in many studies,
but they performed it only in the forward direction
\cite{Xiong_arXiv2017,Xiong_ICASSP2018,Xiong_EMNLP2018,Parthasarathy_arXiv2019,Irie_ASRU2019,Zmolikova_CHiME6,Sun_ICASSP2021}.
%but it is performed in the forward direction [88-88].
%(in [88]).
In contrast, we perform context carry-over
in both forward and backward directions
(in \cite{Xiong_arXiv2017,Xiong_ICASSP2018},
the authors
%claimed
claim that
they performed
%it
context carry-over
in both directions
with $N$-best rescoring,
but
%however,
they
%did
do
not report the effectiveness of
performing it
%context carry-over
in the
backward direction).
%is not reported).
%
%Also, we perform context carry-over
%in up to eight iterative lattice generation.
%
%%%%%%%%%%%%%%%%%%%%%%%%%%%%%%%%%%%%%%%%%%%%%%%%%%%%%%%%%%%%%%%%%%%%%%%%%%%%%%%
\vspace{-2.25mm}
%\vspace{-2.00mm}
\section{Experiments}
\label{sec_exp}
\vspace{-1.25mm}
%%%%%%%%%%%%%%%%%%%%%%%%%%%%%%%%%%%%%%%%%%%%%%%%%%%%%%%%%%%%%%%%%%%%%%%%%%%%%%%
%
To confirm the effectiveness of using a large ensemble
of complementary NLMs on lattice rescoring,
we conducted experiments using the corpus of spontaneous Japanese
(CSJ) \cite{Maekawa_SSPR2003},
which is a large-scale lecture speech corpus.
We performed ASR using the Kaldi hybrid ASR system
\cite{Povey_ASRU2011}
%.
%%%%%based on
%%%%%%with
%%%%%its CSJ ASR recipe.
%%%%%%
%%%%%We used PyTorch
%%%%%\cite{Paszke_NeurIPS2019}
%%%%%%for all the NN modeling in this study.
%%%%%to train the NLMs.
and trained the NLMs using PyTorch
\cite{Paszke_NeurIPS2019,PyTorchExamples}.
%
%******************************************************************************
\vspace{-5.75mm}
%\vspace{-2.00mm}
\subsection{Experimental settings}
\label{ssec_exp_set}
\vspace{-0.00mm}
%******************************************************************************
%
%The CSJ training data consists of
%516 hours,
%403k utterances (sentences),
%and 7.7M words.
%
%The development data consists of
%6.5 hours,
%4k utterances,
%%(sentences)
%and 96k words,
%which was used for any hyperparameter tuning in this study.
%%
%Details of the three evaluation datasets are shown in Table~\ref{tab_data}.
%Table~\ref{tab_data} shows Details of the three evaluation datasets.
%
Details of the CSJ training,
development,
and evaluation datasets
are shown in Table~\ref{tab_data}
(the
original
Kaldi
CSJ recipe has three evaluation datasets,
but we merged them for simplicity).
Using the training data,
we trained a time delay NN-based acoustic model
\cite{Peddinti_Interspeech2015}
and a trigram LM
\cite{Jelinek_MITPress1998,Kneser_ICASSP1995}.
The vocabulary size was set at 44k
%(the words that appear only one time in the training data were mapped to the
%unknown word).
(words that appear only one time in the training data were mapped to
the
%an
unknown word).
%
%%%%%Perplexities of this trigram LM for the dev/eval data are shown in
%%%%%Table~\ref{tab_ppl}.
%
Using the acoustic model and the trigram LM,
we performed
%weighted finite-state transducer-based
one-pass decoding
\cite{Allauzen_CIAA2007} for the dev/eval data
and obtained lattices for all the dev/eval utterances.
%as ASR hypotheses.

%%%%%We used the
%%%%%%word
%%%%%PyTorch
%%%%%NLM training tool
%%%%%%included in PyTorch examples
%%%%%\cite{PyTorchExamples}
%%%%%to train NLMs.
%
Using the training data and
%the training tool,
the PyTorch NLM training tool
\cite{PyTorchExamples},
we trained
%a forward LSTMLM and a forward Transformer-LM,
forward/backward LSTM/Transformer-LMs
%,
%which have the structures shown
having the structures shown
in Table~\ref{tab_nlms}.
%,
%for 80 epochs
%(see the default hyperparameters
%of the tool \cite{PyTorchExamples}
%for detailed training settings).
%(see the tool \cite{PyTorchExamples}
%default hyperparameters
%for detailed training settings).
%
%%%%%Their training settings are
%%%%%%basically
%%%%%the same as the tool default settings
%%%%%%\cite{PyTorchExamples}
%%%%%excepting a small initial learning rate (5.0)
%%%%%and a large backpropagation through time (BPTT) length (256)
%%%%%for the Transformer-LM.
%
The final models (training epochs) were selected based on their
perplexities for the development data.
%
%%%%%Then,
%%%%%we trained backward versions of the above two NLMs
%%%%%using the reversed training data.
%
%%%%%Their structures and training settings were the same as those
%%%%%of the forward NLMs.
%
For the backward versions of the NLMs,
we used the reversed training data.
%
%%%%%Perplexities of the above four NLMs for the dev/eval data are
%%%%%shown in Table~\ref{tab_ppl}.
%
%%%%%Parameters of the above four NLMs were initialized with seed 1.
%%%%%%
%%%%%We also trained seed 2 versions of the four NLMs,
%%%%%i.e., we trained eight complementary NLMs in total.
%%%%%%(i.e., eight NLMs in total).
%%%%%%
%%%%%The seed 2 NLMs showed similar dev/eval data perplexities to those of
%%%%%the seed 1 NLMs shown in Table~\ref{tab_ppl}.
%
We trained two versions of each NLM by changing the
random seed
(i.e., 1 or 2)
%(1 or 2)
for parameter initialization,
i.e., we trained eight NLMs in total.
Perplexities
%of
obtained with
the trigram LM and the four seed 1 NLMs
for the dev/eval data are shown in Table~\ref{tab_ppl}
(the seed 2 NLMs show similar perplexities
as those of the seed 1 NLMs).

%Note that
The training tool
\cite{PyTorchExamples}
concatenates all the training sentences
and then makes a batch data filled with
%the batch size times the
%BPTT (backpropagation through time) length
%backpropagation through time (BPTT) length
%backpropagation through time length
%(the batch size)
%{\tt the batch size}
%``the batch size''
%$\times$
%times
%(the backpropagation through time length)
%``the backpropagation through time length''
the batch size times
%the BPTT length
the backpropagation through time (BPTT) length
of words by splitting the concatenated
long
sentence.
%
%This is because,
%with this batch making strategy,
%zero-padding can be avoided,
%and thus the efficiency of the GPU usage can be greatly improved
%\cite{Irie_ASRU2019}.
This batch-making strategy aims to avoid zero-padding
and maximize GPU usage \cite{Irie_ASRU2019}.
%,
%and,
%with this strategy,
With this strategy,
context carry-over is
%naturally
performed
naturally
in NLM training.
%
%%%This means that the NLMs were trained using the word sequences
%%%that 
%%%%continue
%%%extend
%%%across sentence (utterance) boundaries
%%%and this
%%%training
%%%setting did not match
%%%both utterance-level rescoring (without using contextual information)
%%%and lecture-level rescoring (using contextual information).
%%%%
%%%Solving this mismatch is
%%%one of
%%%our future work
%%%\cite{Parthasarathy_arXiv2019,Irie_ASRU2019,Sun_ICASSP2021}.
%
%Note also that
Furthermore,
the sizes of the trained NLMs are
%appropriately large,
large enough,
%%%%%.
i.e., NLMs with larger sizes started to overfit
the training data.
%%%%%%
%%%%%We observed the
%%%%%trend of
%%%%%overfitting when we trained NLMs
%%%%%%with sizes
%%%%%that have sizes
%%%%%%that are
%%%%%slightly larger
%%%%%than those shown in Table~\ref{tab_nlms}.

Using the trained NLMs
(up to eight NLMs),
%Using the trained (up to eight) NLMs
%%%%%and our
%%%%%%python-based
%%%%%PyTorch-based
%%%%%rescoring tool,
we performed rescoring
on the lattices of all the dev/eval utterances
through iterative lattice generation
%(NLM combination, Section~\ref{ssec_iter})
(Section~\ref{ssec_iter})
%as described in Section~\ref{ssec_iter}.
with and without performing context carry-over
(Section~\ref{ssec_cco}).
From the results of
%the
preliminary experiments,
when we performed context carry-over with
%the
Transformer-LMs,
we set the context length $J$ at 1
%the context length $J$ was set at 1
(see Fig.~\ref{fig_cco}),
since we could not obtain any further performance improvement
by setting $J\!\geq\!2$.
We applied 5-gram approximation for merging hypotheses at a lattice node
and set the maximum number of hypotheses ($k$) stored at a node at 10.
With
%these search settings,
this search setting,
the structure of a generated lattice
at an iteration
%can be changed
can change
from that of the lattice at the previous iteration
\cite{Sundermeyer_IEEEACMTASLP2015,Liu_IEEEACMTASLP2016,Kumar_ASRU2017,Xu_ICASSP2018}.
For further comparison,
%as described in Section~\ref{ssec_others},
%we also performed simultaneous NLM combination
we also performed lattice combination
\cite{Li_IS2002,Xu_ICASSP2010}
and 100-best rescoring
(Section~\ref{ssec_others}).
The 100-best lists were extracted from the lattices.
Hereafter,
we refer to the forward/backward LSTMLMs
trained with seed $x$ as LF$x$ and LB$x$,
and similarly,
we refer to the Transformer-LMs as TF$x$ and TB$x$.
%
%------------------------------------------------------------------------------
\begin{table}[t]
\caption{\ftblskip Details of the CSJ train/dev/eval datasets.}
\vspace{1.0mm}
\label{tab_data}
\begin{center}
\begin{tabular}{lccccc}
\hline
      & Hours & \#lecs & \#utts & \#words & OOV rate \\ \hline
Train &   516 &   3176 &   403k &    7.7M & 0.37$\%$ \\%\hline
Dev   &   6.5 &     39 &   4000 &    9.6k & 1.00$\%$ \\%\hline
Eval  &   5.1 &     30 &   3949 &    7.4k & 0.86$\%$ \\ \hline
\end{tabular}
\end{center}
\vspace{-3.75mm}
%\end{table}
%------------------------------------------------------------------------------
%\begin{table}[t]
%\caption{\ftblskip Structures of the LSTMLM and Transformer-LM.}
\caption{\ftblskip Structures of an LSTMLM and a Transformer-LM.}
\vspace{-2.50mm}
\label{tab_nlms}
\begin{center}
\begin{tabular}{lcc} \hline
                                          &  LSTM & Transformer \\ \hline
Embedding dimensions                      &  1000 &         256 \\
Positional encoding                       &   --- &  Sinusoidal \\
Number of heads                           &   --- &           8 \\
Number of hidden nodes                    &  1000 &        2000 \\
Number of layers                          &     2 &           8 \\
Softmax (vocabulary) size \hspace{3.25mm} & 43720 &       43720 \\
%Number of parameters [M]\,\,\,\,\,\,     & 103.5 &        32.8 \\
\hline
\end{tabular}
\end{center}
\vspace{-6.50mm}
\end{table}
%------------------------------------------------------------------------------
%
%******************************************************************************
\vspace{-2.00mm}
%\subsection{Basic experimental results}
\subsection{Effects of forward/backward NLMs and context carry-over}
\vspace{-0.25mm}
%******************************************************************************
%
Table~\ref{tab_res} shows
%basic
lattice rescoring results obtained with
LF1, LB1, TF1, and TB1.
First,
we can confirm that they (models 1 to 4) steadily reduce the WERs
from the
%strong
ASR 1-best (trigram LM) baseline.
The LSTMLMs show slightly better performance than the Transformer-LMs.
Second,
we can confirm that,
by using
%the
contextual information
%by performing context carry-over
(models 5 to 8),
the WERs can be further reduced.
We can confirm the effect of using contextual information
not only in the forward direction
\cite{Xiong_arXiv2017,Xiong_ICASSP2018,Xiong_EMNLP2018,Parthasarathy_arXiv2019,Irie_ASRU2019,Zmolikova_CHiME6,Sun_ICASSP2021}
%,
but also in the backward direction.
%we can confirm the effect of using the contextual information.
%(this is).
%
%The effect
This effect
%of using contextual information
is larger for
the LSTMLMs than the Transformer-LMs.
This is because,
in contrast to the Transformer-LMs that can use only
%the limited
a limited
%(shrunk)
length of context,
the LSTMLMs can use the
%unlimited length
whole length
of context
as described in Section~\ref{ssec_cco}.
%
%%%%%For Transformer-LMs,
%%%%%we investigated using contexts of lengths ($J$) 1 and 2
%%%%%%(see Section~\ref{ssec_cco}).
%%%%%(see Eq.~(\ref{eq_shrink})),
%%%%%but
%%%%%%However,
%%%%%we cannot
%%%%%%observe
%%%%%confirm
%%%%%the effect of
%%%%%%the different context lengths
%%%%%%the context length differences
%%%%%%on the WER reductions.
%%%%%using the larger context length
%%%%%($J\!=\!1$ is slightly better than $J\!=\!2$).
%%%%%%(context length $J\!=\!1$ corresponds to around 19 words
%%%%%%as shown in Table~\ref{tab_data}).
%%%%%%
%%%%%%Also, we
%%%%%We also cannot obtain any further WER reductions
%%%%%by setting $J\!\geq\!3$.

Third,
we can confirm that,
by combining the forward and backward NLMs iteratively
(models 9 and 10),
the WERs can be greatly reduced
compared
%with when
to the case of
using them individually (models 1 to 4).
%
%This is clearly the effect of the complementarity of the combined NLMs
This is
%clearly
the effect of combining the complementary NLMs
\cite{Xiong_arXiv2017,Xiong_ICASSP2018,Irie_IS2018,Kanda_CHiME5,Arora_CHiME6}.
The effect of using both the forward and backward NLMs is
slightly larger for the Transformer-LMs than for the LSTMLMs.
%
%At last,
Finally,
%for the LSTMLMs,
we can confirm that,
%by performing two iterative rescoring using contextual information
by combining the two LSTMLMs
%iteratively
%with
using
%the
contextual information
(model 11),
the WERs can be further reduced.
This result indicates the complementarity of 
combining
%the
NLMs
and using
%the
contextual information.
In contrast,
%to the LSTMLMs,
%for the Transformer-LMs (models 14 and 15),
%with the Transformer-LMs (models 14 and 15),
with the two Transformer-LMs (model 12),
%the effect of using contextual information
%with NLM combination
%is
%negligible.
the effect is
%very
small.
We need to further investigate
%how to effectively
%use contextual information for the Transformer-LMs.
a
%framework
method
%for effectively using
%for the effective use of
%contextual information with the Transformer-LMs
for effectively carrying over
contextual information with the Transformer-LMs
\cite{Irie_ASRU2019,Sun_ICASSP2021}.
%
%------------------------------------------------------------------------------
\begin{table}[t]
\caption{\ftblskip Dev/Eval data perplexities obtained with the 3g LM and the forward/backward LSTM/Transformer-LMs trained with seed 1.}
\vspace{-2.25mm}
\label{tab_ppl}
\begin{center}
\begin{tabular}{lcccccc}
\hline
Data\,{\textbackslash}\,Model & 3-gram &  LF1 &  LB1 &  TF1 &  TB1 \\ \hline
Dev                           &   71.2 & 31.6 & 31.2 & 30.3 & 29.2 \\%\hline
Eval                          &   70.3 & 34.8 & 34.2 & 33.1 & 32.1 \\ \hline
\end{tabular}
\end{center}
\vspace{-3.50mm}
\caption{\ftblskip Lattice rescoring results in WER [$\%$] obtained with the forward/backward LSTM/Transformer-LMs trained with seed 1. Asterisks $\ast$ indicate the experimental settings that have not been investigated in the previous studies (Section~\ref{sec_rel}).}
\vspace{-2.55mm}
\label{tab_res}
\begin{center}
%\begin{tabular}{@{}rllcc@{}}
\begin{tabular}{rlccc}
\hline
%&
%&
%& \multicolumn{2}{c}{WER [\%]} \\
No.
& Model
%& Context ($J$)
& Context
& Dev
& Eval \\
\hline
 0. & ASR 1-best (3g)                 & No      & 7.7 & 9.0 \\
\hline
 1. & LF1                             & No      & 6.5 & 7.6 \\
 2. & LB1                             & No      & 6.5 & 7.6 \\
 3. & TF1                             & No      & 6.6 & 7.7 \\
 4. & TB1${}^\ast$                    & No      & 6.6 & 7.9 \\
\hline
 5. & LF1                             & Yes     & 6.2 & 7.3 \\
 6. & LB1${}^\ast$                    & Yes     & 6.2 & 7.4 \\
 7. & TF1                             & Yes     & 6.5 & 7.6 \\
% 7.& TF1                             & Yes (1) & 6.5 & 7.6 \\
% 8.& TF1                             & Yes (2) & 6.5 & 7.6 \\
 8. & TB1${}^\ast$                    & Yes     & 6.6 & 7.8 \\
% 9.& TB1                             & Yes (1) & 6.6 & 7.8 \\
%10.& TB1                             & Yes (2) & 6.7 & 7.8 \\
\hline
 9. & LF1 $\rightarrow$ LB1           & No      & 6.3 & 7.3 \\
10. & TF1 $\rightarrow$ TB1${}^\ast$  & No      & 6.4 & 7.4 \\
%11.& LF1 $\rightarrow$ LB1           & No      & 6.3 & 7.3 \\
%12.& TF1 $\rightarrow$ TB1           & No      & 6.4 & 7.4 \\
\hline
%13.& LF1 $\rightarrow$ LB1           & Yes     & \textbf{5.9} & \textbf{7.0} \\
%14.& TF1 $\rightarrow$ TB1           & Yes (1) & 6.3 & 7.3 \\
%15.& TF1 $\rightarrow$ TB1           & Yes (2) & 6.3 & 7.4 \\
11. & LF1 $\rightarrow$ LB1${}^\ast$  & Yes     & \textbf{5.9} & \textbf{7.0} \\
12. & TF1 $\rightarrow$ TB1${}^\ast$  & Yes     & 6.3 & 7.3 \\
\hline
\end{tabular}
\end{center}
%\vspace{-4.30mm}
\vspace{-5.00mm}
\end{table}
%------------------------------------------------------------------------------
%
%******************************************************************************
%\vspace{-1.75mm}
\vspace{-2.50mm}
%\subsection{Experimental results of combining up to eight NLMs}
\subsection{Effects of combining up to eight NLMs}
\label{ssec_res_iter}
\vspace{-0.35mm}
%******************************************************************************
%
We
%iteratively
combined up to the eight
%complementary
NLMs
iteratively
in the order of
%as,
LF1
$\!\rightarrow\!$ LB1
$\!\rightarrow\!$ TF1
$\!\rightarrow\!$ TB1
$\!\rightarrow\!$ LF2
$\!\rightarrow\!$ LB2
$\!\rightarrow\!$ TF2
$\!\rightarrow\!$ TB2
%,
%,
%with the weighting scheme defined in Eq.~(\ref{eq_beta}).
with the procedure described in Section~\ref{ssec_iter}.
%
%We also used contextual information.
We also performed context carry-over.
%
%%%%%For the Transformer-LMs,
%%%%%%we used contexts of length 1 ($J\!=\!1$ in Eq.~(\ref{eq_cco})).
%%%%%we set the context length $J$ in Eq.~(\ref{eq_shrink}) at 1.
%
With this order,
we
%planed to
aimed
%at
to
first
%reducing
reduce
the WERs largely by using
LF1 $\!\rightarrow\!$ LB1
(models 9 and 11 in Table~\ref{tab_res})
and then
%further reducing
to further reduce
%them
the WERs
%by
using
TF1 $\!\rightarrow\!$ TB1
(models 10 and 12),
%that are complementary with LF1 $\!\rightarrow\!$ LB1.
which are complementary with LF1 $\!\rightarrow\!$ LB1.

Figure~\ref{fig_res}
shows experimental results for the evaluation data.
We can confirm that,
thanks to the complementarity of the eight NLMs,
the WERs can be gradually reduced.
Even at the later iterations
(e.g., at the sixth and seventh iterations),
the WERs
%the WER
can be reduced.
We can confirm again the effect of using contextual information.
We
%can
finally obtained a $6.8\%$ WER,
which corresponds to
%a 2.2\% absolute (24.4\% relative) WER reduction
a $24.4\%$ relative WER reduction
from the
%strong
ASR
%(trigram LM)
1-best baseline
%(9.0\% WER).
of $9.0\%$ WER.
%
%Note that we
We also investigated other NLM combination orders
(e.g.,
TF1
$\!\rightarrow\!$ TB1
$\!\rightarrow\!$ LF1
$\!\rightarrow\!$ LB1
$\!\rightarrow\!$ TF2
$\!\rightarrow\!$ TB2
$\!\rightarrow\!$ LF2
$\!\rightarrow\!$ LB2),
%and
%LF1
%$\!\rightarrow\!$ TF1
%$\!\rightarrow\!$ LB1
%$\!\rightarrow\!$ TB1
%$\!\rightarrow\!$ LF2
%$\!\rightarrow\!$ TF2
%$\!\rightarrow\!$ LB2
%$\!\rightarrow\!$ TB2,
but the current order still performed slightly better than these other orders.
%however,
%the current order still performed slightly better than these other orders.
%
%******************************************************************************
%\vspace{-1.75mm}
\vspace{-2.50mm}
\subsection{Comparison with other rescoring methods}
\vspace{-0.35mm}
%******************************************************************************
%
We combined the eight NLMs simultaneously
with lattice combination
\cite{Li_IS2002,Xu_ICASSP2010}.
%and,
%
With the rich
(but slow)
%(slow)
search setting
described in Section~\ref{ssec_exp_set}
(i.e., 5-gram approximation with $k\!=\!10$),
we obtained the same WERs
as those
of the above-described iterative
%lattice generation
%(NLM combination)
NLM combination
as shown in Table~\ref{tab_comp}.
In contrast,
the iterative combination shows slightly lower WERs compared with the
simultaneous combination
when we perform the fast search,
i.e., 0-gram approximation with $k\!=\!1$
(in this setting,
all hypotheses reaching a lattice node are merged,
and thus the
lattice structures are kept throughout rescoring processing
\cite{Sundermeyer_IEEEACMTASLP2015,Liu_IEEEACMTASLP2016,Kumar_ASRU2017,Xu_ICASSP2018}).
%,
%the iterative combination shows slightly lower WERs compared with the
%simultaneous combination.
%
From this result,
we can confirm that
the iterative (gradual) language score refinement
(Section~\ref{ssec_iter})
%will have the advantage
%would have the potential
would have an advantage
%would have the advantage
%have the potential
%to
in
%make
%realize
achieving
stable
%the
rescoring
%processing
with the fast (but unstable) search setting
over the language score refinement that is always performed
%based on
%using
with
the $n$-gram LM scores
(Section~\ref{ssec_others}).

\begin{figure}[t]
  \centering
  \includegraphics[width=1.000\linewidth]{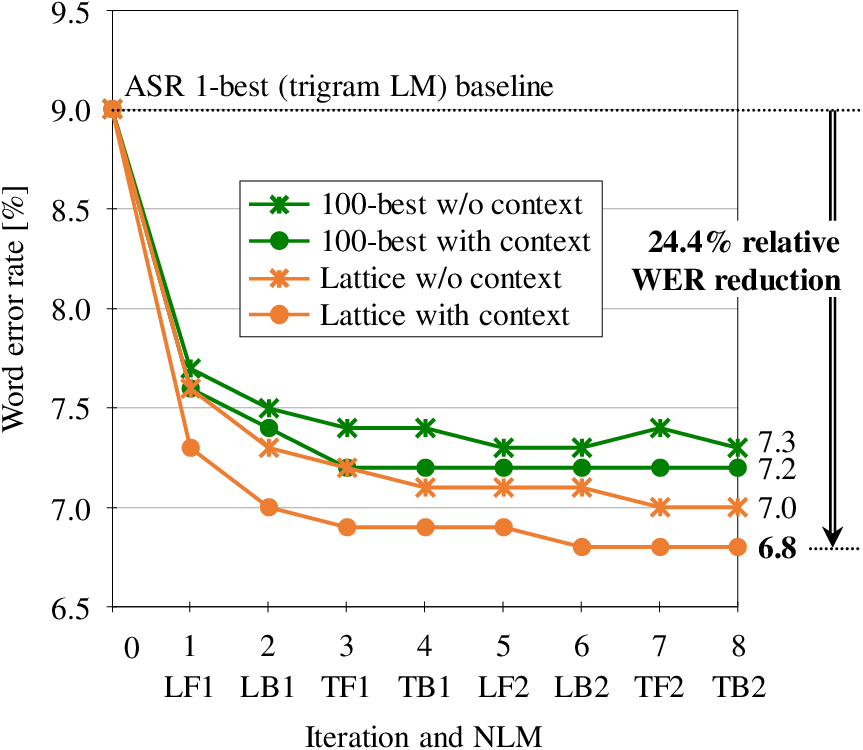}
  \vspace{-5.00mm}
  \caption{\ftblskip Lattice/100-best rescoring results for the evaluation data obtained by combining up to the eight NLMs.}
  \label{fig_res}
  \vspace{-3.25mm}
\end{figure}

We also performed 100-best rescoring by combining up to the eight NLMs
iteratively with
%the above-described order.
the order described above.
The search space of the 100-best lists is greatly limited compared
with that of lattices.
%and thus,
Consequently,
as shown in Fig.~\ref{fig_res},
with 100-best rescoring,
the WER reduction tends to saturate at the earlier iterations
(e.g., the third iteration when using the contextual information).
As a result,
the best WERs achieved with 100-best rescoring
%are
remain
higher
%than those of lattice rescoring.
than those obtained with lattice rescoring.
%
%From this comparison result,
From these comparison results,
we can
confirm the advantage of lattice rescoring over $N$-best rescoring
when using a large ensemble of NLMs.
%
%------------------------------------------------------------------------------
%------------------------------------------------------------------------------
%\begin{table}[t]
%\caption{\ftblskip Lattice rescoring results in WER [$\%$] obtained with the forward/backward LSTM/Transformer-LMs trained with seed 1.}
%%\vspace{1.0mm}
%\label{tab_res2}
%\begin{center}
%\begin{tabular}{rlcccc}
%\hline
%&
%& \multicolumn{2}{c}{Lattice}
%& \multicolumn{2}{c}{100-best} \\
%Iter.
%& Model\,{\textbackslash}\,Context
%& No & Yes & No & Yes \\
%\hline
%   0. & ASR 1-best (3g) & 9.0 &         9.0  & 9.0 & 9.0 \\
%   1. & LF1             & 7.6 & \textbf{7.3} & 7.7 & 7.6 \\
%   2. & LB1             & 7.3 & \textbf{7.0} & 7.5 & 7.4 \\
%   3. & TF1             & 7.2 & \textbf{6.9} & 7.4 & 7.2 \\
%   4. & TB1             & 7.2 & \textbf{6.9} & 7.4 & 7.2 \\
%   5. & LF2             & 7.1 & \textbf{6.9} & 7.3 & 7.2 \\
%   6. & LB2             & 7.0 & \textbf{6.8} & 7.3 & 7.2 \\
%   7. & TF2             & 7.1 & \textbf{6.8} & 7.4 & 7.2 \\
%   8. & TB2             & 7.1 & \textbf{6.8} & 7.3 & 7.2 \\
%\hline
%\end{tabular}
%\end{center}
%\end{table}
%------------------------------------------------------------------------------
\begin{table}[t]
\caption{\ftblskip Comparison of the iterative and simultaneous NLM combinations with the rich and fast search settings for the evaluation data.}
\vspace{-2.75mm}
\label{tab_comp}
\begin{center}
\begin{tabular}{lcccc}
\hline
& \multicolumn{2}{c}{Iterative} & \multicolumn{2}{c}{Simul.} \\
Search setting\,{\textbackslash}\,Context & No & Yes & No & Yes \\
\hline
Rich (5-gram approx., $k=10$) &         7.0  &         6.8  & 7.0 & 6.8 \\
Fast \hspace{0.75mm}(0-gram approx., $k=1$)  & \textbf{7.3} & \textbf{7.0} & 7.5 & 7.1 \\
\hline
\end{tabular}
\end{center}
%\vspace{-6.25mm}
\vspace{-5.20mm}
\end{table}
%------------------------------------------------------------------------------
%
%%%%%%%%%%%%%%%%%%%%%%%%%%%%%%%%%%%%%%%%%%%%%%%%%%%%%%%%%%%%%%%%%%%%%%%%%%%%%%%
\vspace{-0.30mm}
\section{Conclusion and future work}
\vspace{-0.20mm}
%%%%%%%%%%%%%%%%%%%%%%%%%%%%%%%%%%%%%%%%%%%%%%%%%%%%%%%%%%%%%%%%%%%%%%%%%%%%%%%
%%%%%%
%%%%%We
%%%%%%have
%%%%%proposed a lattice rescoring method
%%%%%%that is
%%%%%based on iterative lattice generation
%%%%%using neural LMs.
%%%%%%
%%%%%We experimentally confirmed that the WERs can be gradually reduced
%%%%%along with the iteration increases thanks to the complementarity of the LMs
%%%%%and the WERs can be further reduced by using contextual information
%%%%%across lattices (utterances) for a long speech recording.
%
We experimentally confirmed the effectiveness of using
a large ensemble of complementary NLMs on lattice rescoring.
%
%We believe that 
%the experimental results
The experimental results
and findings
obtained
%in
through
this study are very informative
%since
because
we conducted
%various
a variety of
experiments including
experimental settings that
%were not investigated
%have not yet been investigated
have not been investigated
in the previous studies.
Future work will include the use of more advanced NLMs
%,
%e.g., an LSTMLM with attention mechanism
%\cite{Parthasarathy_arXiv2019}
%and a Transformer-LM that includes an LSTM module
%\cite{Sun_ICASSP2021},
%\cite{Parthasarathy_arXiv2019,Sun_ICASSP2021,Dai_CL2019}
\cite{Parthasarathy_arXiv2019,Sun_ICASSP2021},
%and
%developing
%investigating
an investigation into
a method for effectively weighting the NLMs
in the NLM combination
\cite{Krogh_NIPS1994,Watanabe_ICASSP2014},
%.
and comparison/combination with system combination
\cite{Fiscus_ASRU1997}.
%\vfill
%\pagebreak

%\bibliographystyle{IEEEbib}
%\bibliography{strings,refs}
%\baselineskip 9.14pt
\baselineskip 8.97pt
% \bibliographystyle{IEEEtran}
% For more than N authors, use ``et al''.
% Currently, N is set at 6.
\bibliographystyle{IEEEtran_ogawa}
\bibliography{ogawa}

\end{document}